\documentclass[
]{ceurart}

\usepackage{listings}
\usepackage{colortbl}
\usepackage{amsmath}
\usepackage{multirow}
\usepackage[normalem]{ulem}
\useunder{\uline}{\ul}{}

\usepackage{float}
\begin{document}

\copyrightyear{2023}
\copyrightclause{Copyright for this paper by its authors. Use permitted under Creative Commons License Attribution 4.0 International (CC BY 4.0)}

\conference{Forum for Information Retrieval Evaluation, December 15-18, 2023, India}

\title{Exploring Large Language Models for\\Code Explanation}


\author[1]{Paheli Bhattacharya}[%
email=paheli.bhattacharya@bosch.com
]
\cormark[1]
\fnmark[1]
\author[1]{Manojit Chakraborty}
\fnmark[1]
\author[1]{Kartheek N S N Palepu}
\author[1]{Vikas Pandey}
\author[2]{Ishan Dindorkar}
\author[2]{Rakesh Rajpurohit}
\author[1]{Rishabh Gupta}

\address[1]{Bosch Research and Technology Centre, Bangalore, India}
\address[2]{Bosch Global Software Technologies, Bangalore, India}

\cortext[1]{Corresponding author.}
\fntext[1]{Equal Contribution}

\begin{abstract}
Automating code documentation through explanatory text can prove highly beneficial in code understanding. Large Language Models (LLMs) have made remarkable strides in Natural Language Processing, especially within software engineering tasks such as code generation and code summarization. This study specifically delves into the task of generating natural-language summaries for code snippets, using various LLMs. The findings indicate that Code LLMs outperform their generic counterparts, and zero-shot methods yield superior results when dealing with datasets with dissimilar distributions between training and testing sets.
\end{abstract}


\maketitle
\vspace{-4mm}
\section{Introduction}
\vspace{-2mm}
Understanding legacy codes in large code repositories is a big challenge in the domain of software engineering. Liang et.al.~\cite{liang2018automatic} showed that only 15.4\% of Java GitHub codes are documented. This makes it difficult and time-consuming for developers to comprehend the underlying functionality~\cite{code_comment_lamner, code_summ}.
Automating the task of code documentation through explanations can therefore prove beneficial.

Large Language Models (LLMs) have brought in a progressive breakthrough in Natural Language Processing, especially in the field of Generative AI. LLMs have been applied in many software engineering tasks~\cite{aisw}, popularly in code generation~\cite{codeup}, code summarization~\cite{Wong_2023} and unit test case generation~\cite{schäfer2023empirical}.

In this paper we focus on the task of code explanation -- generating the intent or summary in natural-language for a given code snippet. We benchmark a suite of LLMs -- both generic LLMs~\cite{touvron2023llama} and Code LLMs~\cite{starcoder} using zero-shot, few-shot and instruction fine-tuning approaches. Extensive experiments on the IRSE dataset~\cite{majumdar2023generative} leads to the following insights: (i)~Code LLMs perform better than generic LLMs for the task. (ii)~Zero-shot approaches achieve better results than few-shot and fine-tuning, where the train and test sets follow dissimilar distribution.
\vspace*{-5mm}
\section{Related Work}
\vspace{-2mm}
\textbf{Code explanation}~\cite{code_explain}, also termed as code summarization~\cite{code_summ, code_summ_dl1} and comment generation~\cite{code_comment_dl1, code_comment_lamner}, is an important problem in the field of software engineering. Traditional approaches ~\cite{trad1,trad2,trad3} as well as deep learning methods ~\cite{code_comment_dl1,code_comment_lamner} have been attempted for this task.

\textbf{Large Language Models} have been successfully employed in a wide variety of natural language generation tasks~\cite{yang2023harnessing}. The zero shot and few shot capabilities of these systems make them highly adaptable to any NLP task. There are several general domain, open source LLMs like LLama-2~\cite{touvron2023llama}, Alpaca~\cite{taori2023alpaca} and Falcon~\cite{falcon}. There are also Code LLMs which have been trained or finetuned on code-specific data (usually source code files, covering 80+ programming languages). The most popular LLMs for code are OpenAI CodeX and Co-pilot. Among the open source models, we have StarCoder~\cite{starcoder}, CodeUp~\cite{codeup}, CodeLlama~\cite{codellama} and Llama-2-Coder~\cite{llama2coder}.

\textbf{Large Language Models} have been used for \textbf{Code explanation} in a few shot setting~\cite{code_summ, code_comment_llm}. Ahmed et.al.~\cite{code_summ} found that giving few shot examples from the same project gives better results than from a different project. Geng et.al.~\cite{code_comment_llm} show that selecting relevant examples in a few shot setting is an important design criteria.

\vspace{-4mm}
\section{Dataset}
\vspace{-2mm}
\label{sec:dataset}
\begin{table}[tb]
\centering
\caption{Example data points from IRSE and conala-train datasets along with their average lengths (\#words).}
\label{tab:dataset}
\resizebox{\linewidth}{!}{
\begin{tabular}{|c|c|l|l|cc|}
\hline
\multirow{2}{*}{\textbf{Dataset}} & \multirow{2}{*}{\textbf{Size}} & \multicolumn{1}{c|}{\multirow{2}{*}{\textbf{Example Code Snippet}}} & \multicolumn{1}{c|}{\multirow{2}{*}{\textbf{Example Code Explanation}}} & \multicolumn{2}{c|}{\textbf{Avg. length}} \\ \cline{5-6} 
 &  & \multicolumn{1}{c|}{} & \multicolumn{1}{c|}{} & \multicolumn{1}{c|}{\textbf{Code}} & \textbf{Comment} \\ \hline
\multirow{2}{*}{\textbf{IRSE}} & \multirow{1}{*}{100} & \begin{tabular}[c]{@{}l@{}}pattern = re.compile('\textbackslash{}\textbackslash{}s+')\\ sentence = re.sub(pattern, "", sentence)\end{tabular} & \begin{tabular}[c]{@{}l@{}}This code snippet uses the re (regular expression) module in Python\\ to define a pattern that matches one or more whitespace characters.\\ It then uses the re.sub() function to remove any occurrences of the\\ pattern from the string variable 'sentence'. The result is a modified\\ version of 'sentence' with all whitespace characters removed.\end{tabular} & \multicolumn{1}{c|}{\multirow{1}{*}{21.18}} & \multirow{1}{*}{84.28} \\ \hline
 
\multirow{1}{*}{\textbf{conala-train}} & \multirow{1}{*}{1666} & re.sub('{[}\textasciicircum{}A-Z{]}', '', s) & remove uppercased characters in string `s` & \multicolumn{1}{c|}{\multirow{1}{*}{13.92}} & \multirow{1}{*}{14.68} \\ \hline
 
\end{tabular}
}
\end{table}
 In this work, we consider a dataset of 100 samples released at the Information Retrieval in Software Engineering (IRSE) track at Forum for Information Retrieval Evaluation (FIRE) 2023~\cite{majumdar2023generative}. Each sample in the dataset is a $(code\ snippet, code\ explanation)$ pair. The explanation is a natural language description that denotes what task the code snippet is performing. We refer to this dataset as "IRSE" in the rest of the paper. Additionally, we use a publicly available conala-train~\cite{conala} dataset as a secondary data source for few-shot and instruction finetuning. This dataset consists of 1666 unique samples of $(code\ snippet, code\ explanation)$ pairs.
 
 Table~\ref{tab:dataset} shows a few examples from both the datasets. It can be observed that while the code snippets in are comparable in length
 (21 and 14 tokens respectively), the code explanations in the IRSE dataset are lengthier (mean length = 84 words) than the ones in the conala-train set (mean length = 15 words). 
\vspace{-4mm}

\section{Evaluation}
\label{sec:eval}
\vspace{-2mm}
The model generated textual descriptions are evaluated with respect to the ground truth explanations using the following measures:\\
\noindent
(i)~\textbf{Token-based}: BLEU~\cite{bleu} score combines precision scores of n-grams (typically up to 4-grams) using weighted geometric mean, with higher weight given to shorter n-grams. BLEU-1, BLEU-2, and BLEU-N (for any integer N) extend the evaluation to unigrams, bigrams, and n-grams of varying lengths, respectively.

\noindent
(ii)~\textbf{Semantics-based}: We use this measure to assess the semantic similarity between the model generated explanation ($m$) and the ground truth explanation ($g$). We project both $m$ and $g$ in a continuous embedding space, $ \overrightarrow{e_m}$ and $ \overrightarrow{e_g}$ respectively using the pretrained CodeBERT~\cite{codebert} model. We then take a cosine similarity between the embeddings $cosine( \overrightarrow{e_m},  \overrightarrow{e_g})$ to get the score.
\vspace{-4mm}

\section{Methodology}
\label{sec:method}
\vspace{-2mm}
\begin{table}[tb]
\caption{Zero-shot prompt templates used for Code Explanation. $\{code\}$ denotes the query code snippet whose explanation needs to be generated.}
\label{tab:prompts}
\centering
\resizebox{\linewidth}{!}{
\begin{tabular}{|c|l|c|}
\hline
\textbf{\#} & \multicolumn{1}{c|}{\textbf{Prompt}} & \textbf{Models} \\ \hline
P1 & \begin{tabular}[c]{@{}l@{}}{[}INST{]} \textless \textgreater\\ You are an expert in Programming. Below is a line of python code that describes a task.\\ Return only one line of summary that appropriately describes the task that the code is\\ performing. You must write only summary without any prefix or suffix explanations. \\ Note: The summary should have minimum 1 words and can have on an average 10 words.\\ \textless \textgreater\\ \{\textit{code}\} {[}/INST{]}\end{tabular} & \begin{tabular}[c]{@{}c@{}}Llama-2-70B-Chat\\ CodeLlama-13B-Instruct\\ CodeUp-13B-Chat\end{tabular} \\ \hline
P2 & \begin{tabular}[c]{@{}l@{}}\#Human: You are a helpful code summarizer. Please describe in simple english the\\ purpose of the following Python code snippet: \{\textit{code}\}\\ \#Assistant:\end{tabular} & \begin{tabular}[c]{@{}c@{}}StarCoder (15.5B)\\ Llama-2-Coder-7B\end{tabular} \\ \hline
\end{tabular}
}
\end{table}

We experiment with 5 LLMs (i)~Generic LLM: Llama-2-70B-Chat model~\cite{touvron2023llama}, which is the largest, open source model available. (ii)~Code LLM -- Llama-2-Coder-7B~\cite{llama2coder}, CodeLlama-13B-Instruct~\cite{codellama}, CodeUp-13B-Chat~\cite{codeup} and StarCoder~\cite{starcoder} (15.5B) models, using the zero-shot, few-shot and instruction fine-tuning strategies, described below:\\
\noindent
~\textbf{(i)~Zero-shot}: In this setting, we directly prompt the LLM to generate output for a particular input code snippet. We experiment with several prompts, some of which are listed in Table~\ref{tab:prompts} as prompts P1 and P2. 
 
Based on the model cards, we provide the prompt template P1 to the Llama-2-70B Chat, CodeLlama-13B-Instruct and CodeUp-13B-Chat models. The template P2 is provided to StarCoder and Llama-2-Coder-7B models.\\
\noindent    
~\textbf{(ii)~Few-shot}: In few shot prompting, we provide a few examples that demonstrate the nature of the task. For the task of code explanation ~\cite{code_summ} suggest using 10 examples in a few-shot setup. Therefore, we provide 10 randomly selected $(code\ snippet, natural\ language\ description)$ pairs selected from the conala-train set (ref. Section~\ref{sec:dataset}).\\
\noindent
~\textbf{(iii)~Instruction Finetuning}: For instruction finetuning of LLMs, we take CodeUp-13B-Chat model~\cite{codeup}. We take each sample from conala-train dataset and generate instruction based training instances using the following format:
    \vspace{1mm}
    \\
    \textit{Below is an instruction that describes a task, paired with an input that provides further context. Write a response that appropriately completes the request.\\
    \#\#\# Instruction : Below is a line of python code that describes a task. Write one line of summary that appropriately describes the task that the code is performing.\\
    \#\#\# Input : $sorted(l, key=lambda x: (-int(x[1]), x[0]))$\\
    \#\#\# Output : Sort a nested list by two elements}

We load the CodeUp-13B-Chat model with 4-bit quantization using QLoRA~\cite{dettmers2023qlora} and bitsandbytes~\cite{dettmers2022optimizers} methods. We then perform parameter-efficient finetuning~(PEFT)~\cite{peft} of the model using the above prepared dataset.
\vspace{-4mm}

\section{Results}
\vspace{-2mm}
Table~\ref{tab:results} shows the performance of the 5 different LLMs over three approaches -- zero-shot, few-shot and zero-shot over the Instruction finetuned model. 
\begin{table}[tb]
\centering
\caption{Performance evaluation of the different LLMs and three approaches for the code explanation task on the IRSE dataset. We report the BLEU and CodeBERT based metrics. }
\label{tab:results}
\resizebox{\linewidth}{!}{
\begin{tabular}{|c|c|ccc|c|}
\hline
\multirow{2}{*}{\textbf{Approach}} & \multirow{2}{*}{\textbf{LLM}} & \multicolumn{3}{c|}{\textbf{Token-based}} & \textbf{Semantics-based} \\ \cline{3-6} 
 &  & \multicolumn{1}{c|}{BLEU1} & \multicolumn{1}{c|}{BLEU2} & BLEUN & CodeBERT \\ \hline
\multirow{5}{*}{Zero Shot} & Llama2-70B-Chat & \multicolumn{1}{c|}{0.019} & \multicolumn{1}{c|}{0.008} & 0.004 & 0.338 \\ \cline{2-6} 
 & CodeLlama-13B-Instruct & \multicolumn{1}{c|}{\textbf{0.189}} & \multicolumn{1}{c|}{0.073} & \textbf{0.036} & \textbf{0.498} \\ \cline{2-6} 
 & CodeUp-13B & \multicolumn{1}{c|}{0.010} & \multicolumn{1}{c|}{0.003} & 0.001 & 0.310 \\ \cline{2-6} 
 & StarCoder-15.5B & \multicolumn{1}{c|}{0.069} & \multicolumn{1}{c|}{0.024} & 0.005 & 0.336 \\ \cline{2-6} 
 & Llama-2-Coder-7B & \multicolumn{1}{c|}{\textbf{0.189}} & \multicolumn{1}{c|}{\textbf{0.075}} & 0.023 & 0.475 \\ \hline
\multirow{5}{*}{Few Shot} & Llama2-70B-Chat & \multicolumn{1}{c|}{0.064} & \multicolumn{1}{c|}{0.024} & 0.012 & 0.424 \\ \cline{2-6} 
 & CodeLlama-13B-Instruct & \multicolumn{1}{c|}{\textbf{0.164}} & \multicolumn{1}{c|}{0.073} & 0.044 & \textbf{0.483} \\ \cline{2-6} 
 & CodeUp-13B & \multicolumn{1}{c|}{0.061} & \multicolumn{1}{c|}{0.023} & 0.011 & 0.416 \\ \cline{2-6} 
 & StarCoder-15.5B & \multicolumn{1}{c|}{0.020} & \multicolumn{1}{c|}{0.006} & 0.002 & 0.347 \\ \cline{2-6} 
 & Llama-2-Coder-7B & \multicolumn{1}{c|}{0.023} & \multicolumn{1}{c|}{\textbf{0.008}} & 0.003 & 0.342 \\ \hline
\begin{tabular}[c]{@{}c@{}}Instruction Finetuning\\ Zero Shot\end{tabular} & CodeUp-13B & \multicolumn{1}{c|}{0.047} & \multicolumn{1}{c|}{0.011} & 0.005 & 0.429 \\ \hline
\end{tabular}
}
\end{table}
CodeLlama-13B-Instruct and Llama-2-Coder-7B have the best zero-shot performance over the other LLMs. Note that although the generic Llama2 model is the largest in size (70B), it has poor performance when compared to the smaller Code LLM models (13B, 7B). This shows that domain specific models perform better than generic ones. 

While the few shot strategy is expected to give better performance than zero-shot, in this study we find that the performance is worse. This is mainly because the few shot examples had been selected from the conala-train set. As discussed in Section~\ref{sec:dataset} and Table~\ref{tab:dataset} the code explanation lengths in the IRSE dataset and the conala-train dataset vary hugely. Since the LLMs see few shot examples from the conala-train, it generates shorter length code explanations for input samples coming from the IRSE dataset. This train-test distribution mismatch causes the models to perform worse in the few shot scenario as compared to the zero-shot.

Similar arguments can be drawn for the Instruction finetuning+Zero shot approach, as the training data comes from the conala-train dataset which is different from the IRSE dataset.
\vspace{-4mm}

\section{Conclusion}
\vspace{-2mm}
In this work we explore the performance of 5 LLMs, both generic and code-specifc, for the task of code explanation. We use zero-shot, few shot and instruction finetuning approaches over the LLMs and assess their performance. We find that Code LLMs perform better than larger generic LLMs. Also, zero-shot prompting works well in the scenario where we do not have enough examples to prompt/finetune the model. 
\bibliography{references}

\begin{thebibliography}{28}
\expandafter\ifx\csname natexlab\endcsname\relax\def\natexlab#1{#1}\fi
\providecommand{\url}[1]{\texttt{#1}}
\providecommand{\href}[2]{#2}
\providecommand{\path}[1]{#1}
\providecommand{\DOIprefix}{doi:}
\providecommand{\ArXivprefix}{arXiv:}
\providecommand{\URLprefix}{URL: }
\providecommand{\Pubmedprefix}{pmid:}
\providecommand{\doi}[1]{\href{http://dx.doi.org/#1}{\path{#1}}}
\providecommand{\Pubmed}[1]{\href{pmid:#1}{\path{#1}}}
\providecommand{\bibinfo}[2]{#2}
\ifx\xfnm\relax \def\xfnm[#1]{\unskip,\space#1}\fi
\bibitem[{Liang and Zhu(2018)}]{liang2018automatic}
\bibinfo{author}{Y.~Liang}, \bibinfo{author}{K.~Zhu},
\newblock \bibinfo{title}{Automatic generation of text descriptive comments for code blocks},
\newblock in: \bibinfo{booktitle}{Proceedings of the AAAI Conference on Artificial Intelligence}, volume~\bibinfo{volume}{32}, \bibinfo{year}{2018}.
\bibitem[{Sharma et~al.(2022)Sharma, Chen, and Fard}]{code_comment_lamner}
\bibinfo{author}{R.~Sharma}, \bibinfo{author}{F.~Chen}, \bibinfo{author}{F.~Fard},
\newblock \bibinfo{title}{Lamner: code comment generation using character language model and named entity recognition},
\newblock in: \bibinfo{booktitle}{Proceedings of the 30th IEEE/ACM International Conference on Program Comprehension}, \bibinfo{year}{2022}, pp. \bibinfo{pages}{48--59}.
\bibitem[{Ahmed and Devanbu(2022)}]{code_summ}
\bibinfo{author}{T.~Ahmed}, \bibinfo{author}{P.~Devanbu},
\newblock \bibinfo{title}{Few-shot training llms for project-specific code-summarization},
\newblock in: \bibinfo{booktitle}{Proceedings of the 37th IEEE/ACM International Conference on Automated Software Engineering}, \bibinfo{year}{2022}, pp. \bibinfo{pages}{1--5}.
\bibitem[{Ozkaya(2023)}]{aisw}
\bibinfo{author}{I.~Ozkaya},
\newblock \bibinfo{title}{Application of large language models to software engineering tasks: Opportunities, risks, and implications},
\newblock \bibinfo{journal}{IEEE Software} \bibinfo{volume}{40} (\bibinfo{year}{2023}) \bibinfo{pages}{4--8}.
\bibitem[{Jiang and Kim(2023)}]{codeup}
\bibinfo{author}{J.~Jiang}, \bibinfo{author}{S.~Kim}, \bibinfo{title}{Codeup: A multilingual code generation llama2 model with parameter-efficient instruction-tuning}, \bibinfo{howpublished}{\url{https://huggingface.co/deepse}}, \bibinfo{year}{2023}.
\bibitem[{Wong et~al.(2023)Wong, Guo, Hang, Ho, and Tan}]{Wong_2023}
\bibinfo{author}{M.-F. Wong}, \bibinfo{author}{S.~Guo}, \bibinfo{author}{C.-N. Hang}, \bibinfo{author}{S.-W. Ho}, \bibinfo{author}{C.-W. Tan},
\newblock \bibinfo{title}{Natural language generation and understanding of big code for {AI}-assisted programming: A review},
\newblock \bibinfo{journal}{Entropy} \bibinfo{volume}{25} (\bibinfo{year}{2023}) \bibinfo{pages}{888}. \URLprefix \url{https://doi.org/10.3390%2Fe25060888}. \DOIprefix\doi{10.3390/e25060888}.
\bibitem[{Schäfer et~al.(2023)Schäfer, Nadi, Eghbali, and Tip}]{schäfer2023empirical}
\bibinfo{author}{M.~Schäfer}, \bibinfo{author}{S.~Nadi}, \bibinfo{author}{A.~Eghbali}, \bibinfo{author}{F.~Tip}, \bibinfo{title}{An empirical evaluation of using large language models for automated unit test generation}, \bibinfo{year}{2023}. \href{http://arxiv.org/abs/2302.06527}{{\tt arXiv:2302.06527}}.
\bibitem[{Touvron et~al.(2023)Touvron, Martin, Stone, Albert, Almahairi, Babaei, Bashlykov, Batra, Bhargava, Bhosale et~al.}]{touvron2023llama}
\bibinfo{author}{H.~Touvron}, \bibinfo{author}{L.~Martin}, \bibinfo{author}{K.~Stone}, \bibinfo{author}{P.~Albert}, \bibinfo{author}{A.~Almahairi}, \bibinfo{author}{Y.~Babaei}, \bibinfo{author}{N.~Bashlykov}, \bibinfo{author}{S.~Batra}, \bibinfo{author}{P.~Bhargava}, \bibinfo{author}{S.~Bhosale}, et~al.,
\newblock \bibinfo{title}{Llama 2: Open foundation and fine-tuned chat models},
\newblock \bibinfo{journal}{arXiv preprint arXiv:2307.09288}  (\bibinfo{year}{2023}). \URLprefix \url{https://huggingface.co/meta-llama/Llama-2-70b-chat-hf}.
\bibitem[{et.al.(2023)}]{starcoder}
\bibinfo{author}{R.~L. et.al.},
\newblock \bibinfo{title}{Starcoder: may the source be with you!},
\newblock \bibinfo{journal}{arXiv preprint arXiv:2305.06161}  (\bibinfo{year}{2023}). \URLprefix \url{https://huggingface.co/bigcode/starcoder}.
\bibitem[{Majumdar et~al.(2023)Majumdar, Paul, Paul, Bandyopadhyay, Dave, Chattopadhyay, Das, Clough, and Majumder}]{majumdar2023generative}
\bibinfo{author}{S.~Majumdar}, \bibinfo{author}{S.~Paul}, \bibinfo{author}{D.~Paul}, \bibinfo{author}{A.~Bandyopadhyay}, \bibinfo{author}{B.~Dave}, \bibinfo{author}{S.~Chattopadhyay}, \bibinfo{author}{P.~P. Das}, \bibinfo{author}{P.~D. Clough}, \bibinfo{author}{P.~Majumder},
\newblock \bibinfo{title}{Generative ai for software metadata: Overview of the information retrieval in software engineering track at fire 2023},
\newblock in: \bibinfo{booktitle}{Forum for Information Retrieval Evaluation, ACM}, \bibinfo{year}{2023}.
\bibitem[{MacNeil et~al.(2023)MacNeil, Tran, Hellas, Kim, Sarsa, Denny, Bernstein, and Leinonen}]{code_explain}
\bibinfo{author}{S.~MacNeil}, \bibinfo{author}{A.~Tran}, \bibinfo{author}{A.~Hellas}, \bibinfo{author}{J.~Kim}, \bibinfo{author}{S.~Sarsa}, \bibinfo{author}{P.~Denny}, \bibinfo{author}{S.~Bernstein}, \bibinfo{author}{J.~Leinonen},
\newblock \bibinfo{title}{Experiences from using code explanations generated by large language models in a web software development e-book},
\newblock in: \bibinfo{booktitle}{Proceedings of the 54th ACM Technical Symposium on Computer Science Education V. 1}, \bibinfo{year}{2023}, pp. \bibinfo{pages}{931--937}.
\bibitem[{Iyer et~al.(2016)Iyer, Konstas, Cheung, and Zettlemoyer}]{code_summ_dl1}
\bibinfo{author}{S.~Iyer}, \bibinfo{author}{I.~Konstas}, \bibinfo{author}{A.~Cheung}, \bibinfo{author}{L.~Zettlemoyer},
\newblock \bibinfo{title}{Summarizing source code using a neural attention model},
\newblock in: \bibinfo{booktitle}{54th Annual Meeting of the Association for Computational Linguistics 2016}, \bibinfo{organization}{Association for Computational Linguistics}, \bibinfo{year}{2016}, pp. \bibinfo{pages}{2073--2083}.
\bibitem[{Hu et~al.(2018)Hu, Li, Xia, Lo, and Jin}]{code_comment_dl1}
\bibinfo{author}{X.~Hu}, \bibinfo{author}{G.~Li}, \bibinfo{author}{X.~Xia}, \bibinfo{author}{D.~Lo}, \bibinfo{author}{Z.~Jin},
\newblock \bibinfo{title}{Deep code comment generation},
\newblock in: \bibinfo{booktitle}{Proceedings of the 26th Conference on Program Comprehension}, \bibinfo{publisher}{Association for Computing Machinery}, \bibinfo{year}{2018}, p. \bibinfo{pages}{200–210}.
\bibitem[{Haiduc et~al.(2010)Haiduc, Aponte, Moreno, and Marcus}]{trad1}
\bibinfo{author}{S.~Haiduc}, \bibinfo{author}{J.~Aponte}, \bibinfo{author}{L.~Moreno}, \bibinfo{author}{A.~Marcus},
\newblock \bibinfo{title}{On the use of automated text summarization techniques for summarizing source code},
\newblock in: \bibinfo{booktitle}{2010 17th Working conference on reverse engineering}, \bibinfo{organization}{IEEE}, \bibinfo{year}{2010}, pp. \bibinfo{pages}{35--44}.
\bibitem[{Eddy et~al.(2013)Eddy, Robinson, Kraft, and Carver}]{trad2}
\bibinfo{author}{B.~P. Eddy}, \bibinfo{author}{J.~A. Robinson}, \bibinfo{author}{N.~A. Kraft}, \bibinfo{author}{J.~C. Carver},
\newblock \bibinfo{title}{Evaluating source code summarization techniques: Replication and expansion},
\newblock in: \bibinfo{booktitle}{2013 21st International Conference on Program Comprehension (ICPC)}, \bibinfo{organization}{IEEE}, \bibinfo{year}{2013}, pp. \bibinfo{pages}{13--22}.
\bibitem[{Moreno et~al.(2013)Moreno, Aponte, Sridhara, Marcus, Pollock, and Vijay-Shanker}]{trad3}
\bibinfo{author}{L.~Moreno}, \bibinfo{author}{J.~Aponte}, \bibinfo{author}{G.~Sridhara}, \bibinfo{author}{A.~Marcus}, \bibinfo{author}{L.~Pollock}, \bibinfo{author}{K.~Vijay-Shanker},
\newblock \bibinfo{title}{Automatic generation of natural language summaries for java classes},
\newblock in: \bibinfo{booktitle}{2013 21st International conference on program comprehension (ICPC)}, \bibinfo{organization}{IEEE}, \bibinfo{year}{2013}, pp. \bibinfo{pages}{23--32}.
\bibitem[{Yang et~al.(2023)Yang, Jin, Tang, Han, Feng, Jiang, Yin, and Hu}]{yang2023harnessing}
\bibinfo{author}{J.~Yang}, \bibinfo{author}{H.~Jin}, \bibinfo{author}{R.~Tang}, \bibinfo{author}{X.~Han}, \bibinfo{author}{Q.~Feng}, \bibinfo{author}{H.~Jiang}, \bibinfo{author}{B.~Yin}, \bibinfo{author}{X.~Hu},
\newblock \bibinfo{title}{Harnessing the power of llms in practice: A survey on chatgpt and beyond},
\newblock \bibinfo{journal}{arXiv preprint arXiv:2304.13712}  (\bibinfo{year}{2023}).
\bibitem[{Taori et~al.(2023)Taori, Gulrajani, Zhang, Dubois, Li, Guestrin, Liang, and Hashimoto}]{taori2023alpaca}
\bibinfo{author}{R.~Taori}, \bibinfo{author}{I.~Gulrajani}, \bibinfo{author}{T.~Zhang}, \bibinfo{author}{Y.~Dubois}, \bibinfo{author}{X.~Li}, \bibinfo{author}{C.~Guestrin}, \bibinfo{author}{P.~Liang}, \bibinfo{author}{T.~B. Hashimoto},
\newblock \bibinfo{title}{Alpaca: A strong, replicable instruction-following model},
\newblock \bibinfo{journal}{Stanford Center for Research on Foundation Models. https://crfm. stanford. edu/2023/03/13/alpaca. html} \bibinfo{volume}{3} (\bibinfo{year}{2023}) \bibinfo{pages}{7}.
\bibitem[{Penedo et~al.(2023)Penedo, Malartic, Hesslow, Cojocaru, Cappelli, Alobeidli, Pannier, Almazrouei, and Launay}]{falcon}
\bibinfo{author}{G.~Penedo}, \bibinfo{author}{Q.~Malartic}, \bibinfo{author}{D.~Hesslow}, \bibinfo{author}{R.~Cojocaru}, \bibinfo{author}{A.~Cappelli}, \bibinfo{author}{H.~Alobeidli}, \bibinfo{author}{B.~Pannier}, \bibinfo{author}{E.~Almazrouei}, \bibinfo{author}{J.~Launay},
\newblock \bibinfo{title}{The {R}efined{W}eb dataset for {F}alcon {LLM}: outperforming curated corpora with web data, and web data only},
\newblock \bibinfo{journal}{arXiv preprint arXiv:2306.01116}  (\bibinfo{year}{2023}).
\bibitem[{Rozi{\`e}re et~al.(2023)Rozi{\`e}re, Gehring, Gloeckle, Sootla, Gat, Tan, Adi, Liu, Remez, Rapin et~al.}]{codellama}
\bibinfo{author}{B.~Rozi{\`e}re}, \bibinfo{author}{J.~Gehring}, \bibinfo{author}{F.~Gloeckle}, \bibinfo{author}{S.~Sootla}, \bibinfo{author}{I.~Gat}, \bibinfo{author}{X.~E. Tan}, \bibinfo{author}{Y.~Adi}, \bibinfo{author}{J.~Liu}, \bibinfo{author}{T.~Remez}, \bibinfo{author}{J.~Rapin}, et~al.,
\newblock \bibinfo{title}{Code llama: Open foundation models for code},
\newblock \bibinfo{journal}{arXiv preprint arXiv:2308.12950}  (\bibinfo{year}{2023}). \URLprefix \url{https://huggingface.co/codellama}.
\bibitem[{{Manuel Romero}(2023)}]{llama2coder}
\bibinfo{author}{{Manuel Romero}}, \bibinfo{title}{llama-2-coder-7b (revision d30d193)}, \bibinfo{year}{2023}. \URLprefix \url{https://huggingface.co/mrm8488/llama-2-coder-7b}. \DOIprefix\doi{10.57967/hf/0931}.
\bibitem[{Geng et~al.(2024)Geng, Wang, Dong, Wang, Li, Jin, Mao, and Liao}]{code_comment_llm}
\bibinfo{author}{M.~Geng}, \bibinfo{author}{S.~Wang}, \bibinfo{author}{D.~Dong}, \bibinfo{author}{H.~Wang}, \bibinfo{author}{G.~Li}, \bibinfo{author}{Z.~Jin}, \bibinfo{author}{X.~Mao}, \bibinfo{author}{X.~Liao},
\newblock \bibinfo{title}{Large language models are few-shot summarizers: Multi-intent comment generation via in-context learning}  (\bibinfo{year}{2024}).
\bibitem[{Yin et~al.(2018)Yin, Deng, Chen, Vasilescu, and Neubig}]{conala}
\bibinfo{author}{P.~Yin}, \bibinfo{author}{B.~Deng}, \bibinfo{author}{E.~Chen}, \bibinfo{author}{B.~Vasilescu}, \bibinfo{author}{G.~Neubig},
\newblock \bibinfo{title}{Learning to mine aligned code and natural language pairs from stack overflow},
\newblock in: \bibinfo{booktitle}{International Conference on Mining Software Repositories}, \bibinfo{publisher}{ACM}, \bibinfo{year}{2018}, pp. \bibinfo{pages}{476--486}. \URLprefix \url{https://conala-corpus.github.io/}.
\bibitem[{Papineni et~al.(2002)Papineni, Roukos, Ward, and Zhu}]{bleu}
\bibinfo{author}{K.~Papineni}, \bibinfo{author}{S.~Roukos}, \bibinfo{author}{T.~Ward}, \bibinfo{author}{W.-J. Zhu},
\newblock \bibinfo{title}{Bleu: A method for automatic evaluation of machine translation},
\newblock \bibinfo{publisher}{Association for Computational Linguistics}, \bibinfo{address}{USA}, \bibinfo{year}{2002}, p. \bibinfo{pages}{311–318}.
\bibitem[{Feng et~al.(2020)Feng, Guo, Tang, Duan, Feng, Gong, Shou, Qin, Liu, Jiang et~al.}]{codebert}
\bibinfo{author}{Z.~Feng}, \bibinfo{author}{D.~Guo}, \bibinfo{author}{D.~Tang}, \bibinfo{author}{N.~Duan}, \bibinfo{author}{X.~Feng}, \bibinfo{author}{M.~Gong}, \bibinfo{author}{L.~Shou}, \bibinfo{author}{B.~Qin}, \bibinfo{author}{T.~Liu}, \bibinfo{author}{D.~Jiang}, et~al.,
\newblock \bibinfo{title}{Codebert: A pre-trained model for programming and natural languages},
\newblock in: \bibinfo{booktitle}{Findings of the Association for Computational Linguistics: EMNLP 2020}, \bibinfo{year}{2020}, pp. \bibinfo{pages}{1536--1547}.
\bibitem[{Dettmers et~al.(2023)Dettmers, Pagnoni, Holtzman, and Zettlemoyer}]{dettmers2023qlora}
\bibinfo{author}{T.~Dettmers}, \bibinfo{author}{A.~Pagnoni}, \bibinfo{author}{A.~Holtzman}, \bibinfo{author}{L.~Zettlemoyer}, \bibinfo{title}{Qlora: Efficient finetuning of quantized llms}, \bibinfo{year}{2023}. \href{http://arxiv.org/abs/2305.14314}{{\tt arXiv:2305.14314}}.
\bibitem[{Dettmers et~al.(2022)Dettmers, Lewis, Shleifer, and Zettlemoyer}]{dettmers2022optimizers}
\bibinfo{author}{T.~Dettmers}, \bibinfo{author}{M.~Lewis}, \bibinfo{author}{S.~Shleifer}, \bibinfo{author}{L.~Zettlemoyer},
\newblock \bibinfo{title}{8-bit optimizers via block-wise quantization},
\newblock \bibinfo{journal}{9th International Conference on Learning Representations, ICLR}  (\bibinfo{year}{2022}).
\bibitem[{Mangrulkar et~al.(2022)Mangrulkar, Gugger, Debut, Belkada, Paul, and Bossan}]{peft}
\bibinfo{author}{S.~Mangrulkar}, \bibinfo{author}{S.~Gugger}, \bibinfo{author}{L.~Debut}, \bibinfo{author}{Y.~Belkada}, \bibinfo{author}{S.~Paul}, \bibinfo{author}{B.~Bossan}, \bibinfo{title}{Peft: State-of-the-art parameter-efficient fine-tuning methods}, \bibinfo{howpublished}{\url{https://github.com/huggingface/peft}}, \bibinfo{year}{2022}.

\end{thebibliography}


\end{document}